\documentclass[twocolumn]{aastex631}
\usepackage{CJK}

\begin{document}
\begin{CJK*}{UTF8}{gbsn}

\title{Detecting Biosignatures in Nearby Rocky Exoplanets using High-Contrast Imaging and Medium-Resolution Spectroscopy with Extremely Large Telescope}

\shorttitle{Detecting Biosignatures in Nearby Rocky Exoplanets}
\shortauthors{Zhang et al.}

\author[0009-0006-9777-3051]{Huihao Zhang}
\affiliation{Department of Astronomy, The Ohio State University, 140 W. 18th Ave., Columbus, OH 43210, USA}

\correspondingauthor{Huihao Zhang}
\email{zhang.12043@buckeyemail.osu.edu}

\author[0000-0002-4361-8885]{Ji Wang (王吉)}
\affiliation{Department of Astronomy, The Ohio State University, 140 W. 18th Ave., Columbus, OH 43210, USA}

\author[0000-0002-4831-0329]{Michael K. Plummer}
\affiliation{Department of Astronomy, The Ohio State University, 140 W. 18th Ave., Columbus, OH 43210, USA}

\begin{abstract}

In the upcoming decades, one of the primary objectives in exoplanet science is to search for habitable planets and signs of extraterrestrial life in the universe. Signs of life can be indicated by thermal-dynamical imbalance in terrestrial planet atmospheres. O$_2$ and CH$_4$ in the modern Earth's atmosphere are such signs, commonly termed biosignatures. These biosignatures in exoplanetary atmospheres can potentially be detectable through high-contrast imaging instruments on future extremely large telescopes (ELTs). To quantify the signal-to-noise ($S/N$) ratio with ELTs, we select up to 10 nearby rocky planets and simulate medium resolution (R $\sim$ 1000) direct imaging of these planets using the Mid-infrared ELT Imager and Spectrograph (ELT/METIS, 3-5.6 $\mu$m) and the High Angular Resolution Monolithic Optical and Near-infrared Integral field spectrograph (ELT/HARMONI, 0.5-2.45 $\mu$m). We calculate the $S/N$ for the detection of biosignatures including CH$_4$, O$_2$, H$_2$O, and CO$_2$. Our results show that GJ 887~b has the highest detection $S/N$ for biosignatures and Proxima Cen b exhibits the only detectable CO$_2$ among the targets for ELT/METIS direct imaging. We also investigate the TRAPPIST-1 system, the archetype of nearby transiting rocky planet systems, and compare the biosignature detection $S/N$ of transit spectroscopy with JWST versus direct spectroscopy with ELT/HARMONI. Our findings indicate JWST is more suitable for detecting and characterizing the atmospheres of transiting planet systems such as TRAPPIST-1 that are relatively further away and have smaller angular separations than more nearby non-transiting planets.

\end{abstract}


\section{Introduction} \label{sec:style_intro}

The question of whether life exists beyond Earth has been debated since Aristotle. We will perhaps have the opportunity to answer this question in the coming decades. The Astro2020 Decadal Survey \citep{national2021decadal}  emphasizes the search for habitable worlds and extraterrestrial life as one of primary goals for the next decade. In addition, a series of high-performance extremely large telescopes (ELTs) are scheduled to come on-line \citep{sanders2013thirty, gilmozzi2007european, roberge2018large}. These ELTs will be vital in searching for the signs of life.

Evidence of life can be reflected in the thermal-dynamical imbalance of planetary atmospheres, which is manifested in certain molecular combinations in planetary atmospheres \citep{schwieterman2018exoplanet, krissansen2018disequilibrium}. These molecular combinations are referred to as biosignature pairs. For example, the thermal-dynamical imbalance leads to the existing combination of O$_2$ and CH$_4$ in the atmosphere of Earth \citep{stueeken2020mission, crouse2021biosignature}.

Transit spectroscopy can be used to detect and characterize the atmosphere of transiting planets \citep{charbonneau2002detection, deming2013infrared,  jwst2023identification}. However, transit spectroscopy has limitations in detecting the atmospheres of nearby planets because of the low transit probability for these exoplanets \citep{malin2023simulated}. Direct imaging can overcome the problem of low transit probability and therefore effectively characterize both transiting and non-transiting planets~\citep{werber2023direct} while obtaining atmospheric and orbital data~\citep{wang2020exoplanet}. 

Direct imaging faces challenges in separating the planetary signal from the noise due to the high contrast and small angular separation between the planet and the host star \citep{mawet2022fiber}. Currently, most of the known exoplanets \citep{marois2008direct,marois2010images, lagrange2010giant, rameau2013confirmation, macintosh2015discovery, chauvin2017discovery} imaged directly are around young host stars ($<100$ Myr) because the planet-star flux ratio (hereinafter called ``contrast``) is higher due to the young planet's preservation of a large amount of heat during planetary formation \citep{fortney2008synthetic,werber2023direct}.

The challenges associated with small angular separation and high contrast in direct imaging can be addressed by future ELTs such as the European Extremely Large Telescope \citep[ELT,][]{gilmozzi2007european}, Thirty Meter Telescope \citep[TMT,][]{sanders2013thirty}, and Giant Magellan Telescope \citep[GMT,][]{johns2006giant}. For example, the coronagraphs in ELT/METIS \citep{brandl2016status} and ELT/HARMONI \citep{jocou2022harmoni} enable the capability of medium-resolution (1000--5000) spectroscopy in the near-infrared (NIR) and mid-infrared (MIR) bands for high-contrast exoplanets \citep{quanz2015direct, brandl2016status,
 wang2017observing, carlomagno2020metis, bowens2021exoplanets}. \citet{werber2023direct} explore the ELT's ability to directly image habitable exoplanets and highlight the MIR band as the most suitable for direct imaging, because planets have more favorable contrast in the MIR \citep{kasper2017near}.

However, exploration of direct imaging by ELTs in the NIR band remains valuable. Compared to MIR, NIR also has strong absorption features of gas \citep{wang2017observing} and offers a low sky background as well as high spatial resolution, which may make it easier to detect exoplanets with small angular separation, such as those around M stars.

\begin{figure*}[t]
\plotone{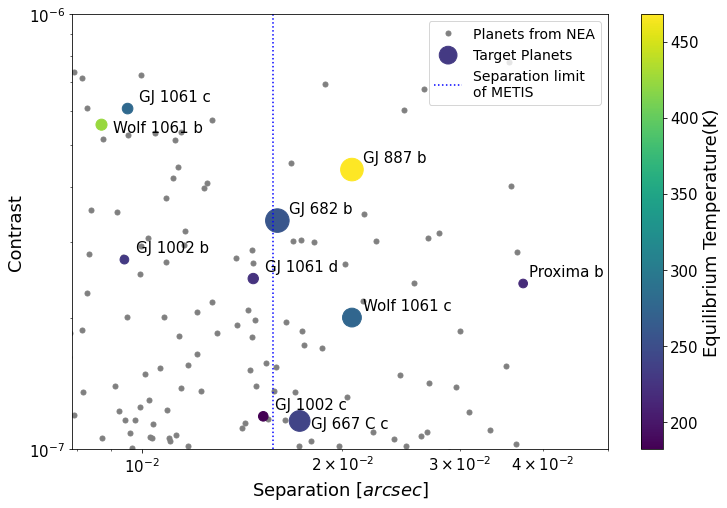}
\caption{We select 10 and 5 candidate planets around nearby stars for ELT/HARMONI and ELT/METIS, respectively. As the spatial resolution of ELT/HARMONI is smaller than the plotting range, we only show spatial resolution for ELT/METIS as the vertical dotted blue line. Planets located to the right of the dotted line are suitable planets for ELT/METIS. All colored points are suitable planets for ELT/HARMONI. The size of the colored point is proportional to the planet radius. The color of the point is based on the equilibrium temperature.  All colored data points are the candidate planets listed in Table \ref{table:general1_4}. Selection criteria can be found in \S \ref{sec:style_sec_planet}. The gray points are other exoplanets from NEA that are less suitable for direct imaging with ELT.
\label{fig:general1_4}}
\end{figure*}

In this study, we use ELT/METIS \citep{brandl2008metis} in 3.0-5.6 $\mu$m and ELT/HARMONI \citep{2016SPIE.9908E..9TK} in 0.5-2.45 $\mu$m as examples to explore and quantify the ability of ELTs to perform medium-resolution, high-contrast direct imaging of nearby known rocky exoplanets, using the methods of \citet{wang2017observing} and \citet{2021ApJ...923..144P}. Our goal is to provide a ranking of the signal-to-noise ($S/N$) ratio of the biosignatures in the atmosphere of rocky exoplanets for ELT/METIS and ELT/HARMONI at different starlight suppression levels.  We will also conduct a comparative study for the TRAPPIST-1 system using both the transit method with JWST and direct imaging with ELT/HARMONI.

In \S \ref{sec:style_sec}, we discuss the selection of candidate planets and instruments. In \S \ref{sec:style_meth}, we discuss the methods used to simulate and quantify the $S/N$ of ELT as well as the selection of planetary template atmospheres. In \S \ref{sec:styleApplication}, we simulate direct imaging of planetary atmospheres using ELT/METIS and ELT/HARMONI. We quantify the $S/N$ of the candidate planets. We then investigate the $S/N$ of ELT/METIS and ELT/HARMONI for the atmosphere of the most ideal candidate planets under multiple instrument contrast levels. In \S \ref{sec:styledisc}, we discuss the $S/N$ of biosignature pairs and compare the detection of biosignatures in exoplanet atmospheres by ELT and JWST. We summarize our results and draw conclusions on our study in \S \ref{sec:stylesum}. We also introduce our publicly available Python code on Zenodo \citep{hhz_code_2023}.

\section{The Selection of Planets and Instruments} \label{sec:style_sec}

\subsection{The Selection of candidate planets} \label{sec:style_sec_planet}

In this section, we select 10 and 5 temperate rocky planets from the NASA Exoplanet Archive (NEA)\footnote{\url{https://exoplanetarchive.ipac.caltech.edu/}} and describe them in Table \ref{table:general1_4}, catering to the ELT/HARMONI and ELT/METIS, respectively. Angular separation and low contrast are key parameters when selecting planetary targets for direct imaging.
We employ NEA and the following selection criteria for ELT/HARMONI and ELT/METIS: (1) declination below 0$^{\circ}$; (2) top 10 planets with separations larger than the resolution for ELT/HARMONI (2.36 mas) at 0.45 $\mu$m; 5 planets with separations larger than the angular resolution for ELT/METIS (15.74 mas) at 3 $\mu$m; (3) contrast above  10$^{-7}$; (4) planets with radii between 1.0 R$_\oplus$ and 2 R$_\oplus$; (5) equilibrium temperature below 647 K. Based on these selection criteria, we select 10 and 5 planets wich the  which are shown in Figure \ref{fig:general1_4}. Below we detail how each of the criteria is being set.

\subsubsection{Selection Criteria} 

\begin{table*}[t]
\setlength{\arrayrulewidth}{0.4mm}
\begin{tabular}{c c c c c c c c c c }
\hline\hline
Planet & M$_{P}$ &R$_{P}$& SMA & D & R$_{\star}$ & $\log(g)_{\star}$& T$_{\star}$ & T$_{eq}$ & $\lambda_{max}$ \\ & (M$_{\oplus}$) &(R$_{\oplus}$) & (AU) & (PC) & (R$_\odot$) & (cm/s$^2$)& (K) & (K) & ($\mu$m) \\
\hline
GJ 1002 c & 1.36 & 1.09 & 0.074 & 4.849 & 0.14 & 5.100 & 3024 & 182 & 2.90\\
GJ 1061 d & 1.64 & 1.15 & 0.054 & 3.673 & 0.16 & 5.160 & 2953 & 224 & 2.80\\
GJ 1002 b & 1.08 & 1.02 & 0.046 & 4.849 & 0.14 & 5.100 & 3024 & 231 & 1.80\\
Wolf 1061 b & 1.91 & 1.20 & 0.038 & 4.306 & 0.31 & 4.900 & 3342 & 424 & 1.66\\
GJ 1061 c & 1.74 & 1.17 & 0.035 & 3.673 & 0.16 & 5.160 & 2953 & 278 & 1.82\\
 $^*$GJ 667 C c & 3.80 & 1.82 & 0.125 & 7.244 & 0.33 & 4.690 & 3350 & 240 & 3.29\\
 $^*$GJ 682 b & 4.40 & 1.98 & 0.080 & 5.007 & 0.30 & 4.930 & 3028 & 259 & 3.04\\
 $^*$GJ 887 b & 4.20 & 1.93 & 0.068 & 3.290 & 0.47 & 4.780 & 3688 & 468 & 3.94\\
 $^*$Proxima b & 1.07 & 1.02 & 0.049 & 1.301 & 0.14 & 5.160 & 2900 & 217 & 7.11\\
 $^*$Wolf 1061 c & 3.41 & 1.71 & 0.089 & 4.306 & 0.31 & 4.900 & 3342 & 275 & 3.94\\[1ex]
\hline
\end{tabular}
\caption{A list of candidate planets that are suitable for direct imaging with ELT. All planets are suitable planets for ELT/HARMONI, the planets with $^*$ are suitable planets for ELT/METIS. M$_{P}$: mass of planet, R$_{P}$: radius of planet, SMA: semi-major axis of planet, D: distance between planet and Earth, R$_{\star}$: radius of star,  $\log(g)_{\star}$: surface gravity of star, T$_{\star}$: surface temperature of star, T$_{eq}$: equilibrium temperature of planet, $\lambda_{max}$: maximum observable wavelength such that planet-star separation is  larger than $\lambda/D$.
\label{table:general1_4}}
\end{table*}

We choose stars with declinations below 0$^{\circ}$ because they are more likely observable at the location of the ELT \citep[southern hemisphere,][]{2016SPIE.9906E..0WT}.

An angular separation limit of 2.36 and 15.74 mas is selected based on the ELT's spatial resolution ($\lambda/D$) computed with a 39.3 m diameter and minimum working wavelength of ELT/HARMONI (0.45 $\mu$m) and ELT/METIS (3\,$\mu$m).

We calculate the planet-star angular separation according to their reported semi-major axis and the distance from Earth.

Based on \citet{wang2020exoplanet}, the planet contrast $\epsilon$ is calculated as follows:

\begin{equation}
\label{eq:single_feature}
\epsilon = \bigg(\frac{R_p^2}{a^2}\bigg)\cdot A
\label{eq:general0_4}
\end{equation}

where $R_p$ is the planetary radius, $a$ is the semi-major axis, and $A$ is the planetary albedo. We adopt a uniform albedo value of $A=0.3$ for all planets. This is the average albedo of Earth \citep{goode2001earthshine}.

For $R_p$, we use values from NEA when available. In cases where the planetary radius is unknown, we utilize the mass-radius relation presented in \citet{chen2016probabilistic} to estimate the planetary radii with mass (or $m\sin i$). For planets detected through the radial velocity method, we assumed an edge-on orbit inclination ($i$) of 90$^\circ$. Planets will have a larger mass and radius if $i \neq$ 90$^\circ$, in this scenario, the $S/N$ of the planet is higher due to the larger radius, which results in a lower planet-star contrast.

We choose planets with radius between 1 and 2 $R_{\oplus}$ because they are more likely to have an Earth-like atmosphere, which is consistent with our choice of template spectra (\S \ref{sec:style_ELT_light_planet}). The imposed contrast limit of 10$^{-7}$ is consistent with \citet{wang2020exoplanet}. 

A 647 K equilibrium temperature limit allows for the possibility of liquid water on the planetary surface \citep{selsis2007habitable}. We use equilibrium temperatures from NEA when available. For cases in which the equilibrium temperature is unknown, we use the following equation to calculate the equilibrium temperature $T_{eq}$:

\begin{equation}
\label{eq:single_feature}
T_{eq} = T_{\star} \sqrt{\frac{R_*}{2a}}(1-A)^{1/4}
\label{eq:general1_4}
\end{equation}

where $T_{\star}$ is the effective temperature of the star and $R_*$ is the radius of the star.

\subsubsection{Results of the Selection} \label{sec:style_seted_p}

We have identified 10 and 5 rocky exoplanets that meet the requirements for ELT/HARMONI and ELT/METIS, respectively, as shown in Figure \ref{table:general1_4}. All of these planets are super-Earths detected through radial velocity. Based on their radii, these planets can be classified into two categories. The first category of planets has radii ranging from 1 R$_{\oplus}$ to 1.20 R$_{\oplus}$. The second category of planets has radii between 1.70 R$_{\oplus}$ and 2 R$_{\oplus}$. 

Planets in the first category with smaller radii approximately equal to Earth's are thought to have lost their massive hydrogen envelopes. The hydrogen in the atmosphere is likely to be lost due to molecular photodissociation \citep{owen2017evaporation} and/or core-powered mass-loss \citep{ginzburg2016super, ginzburg2018core, gupta2019sculpting, gupta2022properties,2023arXiv230800020O}. Planets in this category are more likely to possess an Earth-like atmosphere due to their similar surface gravity. The first category of planets includes GJ 1002 c, GJ 1061 d, Proxima b (or Proxima Cen b), GJ 1002 b, Wolf 1061 b, GJ 1061 c.

For planets within the second category, not only is there a potential for the formation of an Earth-like atmosphere, but there also exists the likelihood of establishing a H-rich atmosphere \citep{miller2008atmospheric}. Unlike the first type of planet, after the formation of the planetary core, due to a larger surface gravity, it is possible to retain hydrogen in the atmosphere. Under favorable conditions, these super-Earths are also deemed habitable \citep{haghighipour2013formation, seager2013exoplanet, huang2022assessment,phillips2022ltt}. The second category of planets consists of GJ 667 C c, GJ 682 b, GJ 887 b, and Wolf 1061 c. We limit the scope of this work to the investigation of Earth-like atmospheres for both categories of planets. 

\begin{figure*}[t]
\plotone{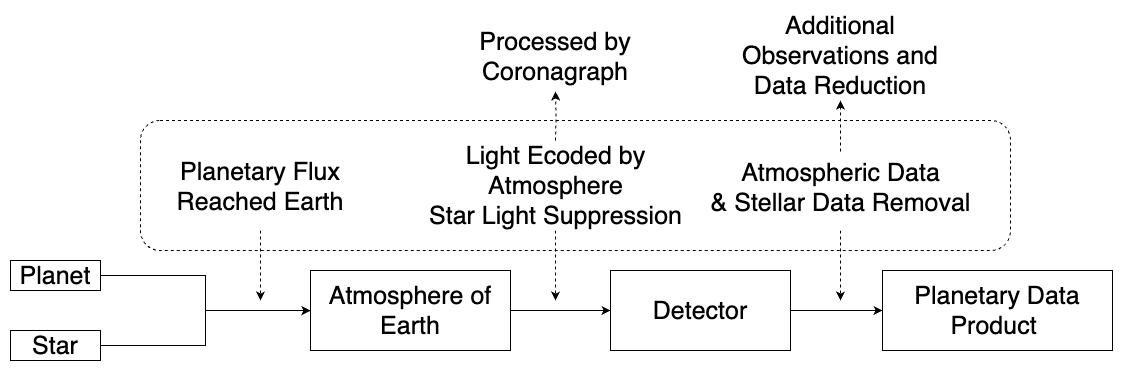}
\caption{Flow chart outlining the simulation of ELT direct imaging applied to exoplanets. Photons from the star and the planet pass through Earth's atmosphere, and processed by the coronagraph, and captured by the detector. After obtaining the data, the effects of the atmosphere and the star are removed through observations of the host star and the positions where no planet signals are received.
\label{fig:flowchart}}
\end{figure*}

\begin{table*}
\setlength{\arrayrulewidth}{0.4mm}
\begin{tabular}{c c c c c c c c c}
\hline\hline
 & Radius & Mass & T${eff}$ & Metallicity & log $g$ & SMA & D & T${eq}$ \\  & & & (K) & (dex) & (cm/s$^2$) & (AU) & (PC) & (K) \\[0.5ex]
\hline
TRAPPIST-1 & 0.1192R$_{\odot}$ & 0.0898M$_{\odot}$ & 2566 & 0.040 & 5.2396 & - & 12.1 & - \\
TRAPPIST-1~e & 0.920R$_{\oplus}$ & 0.692M$_{\oplus}$ & - & - & - & 0.029 & - & 228.5 \\
TRAPPIST-1~h & 0.755R$_{\oplus}$ & 0.326M$_{\oplus}$ & - & - & - & 0.0619 & - & 157.1 \\[1ex]
\hline
\end{tabular}
\caption{TRAPPIST-1 Parameters. All parameters we used are the latest NEA data.\label{table:general1_1_transposed}} 
\end{table*}

\subsection{The Selection of ELT Instruments} \label{sec:style_sec_ins}
We chose to simulate biosignature detectability for ELT/METIS and ELT/HARMONI because both instruments possess the capability of high-contrast imaging with medium-resolution spectroscopy. The choice of medium resolution is due to our method being applicable to medium and low resolutions. The spatial bands covered by ELT/METIS  \citep[3-13 $\mu$m,][]{brandl2018status} and ELT/HARMONI \citep[0.47-2.45 $\mu$m,][]{thatte2016elt} coincide with our desired range of interest (0.5-5.6 $\mu$m). Both instruments are equipped with coronagraphs \citep{carlomagno2020metis, jocou2022harmoni}, possess medium-resolution imaging, and are cable of directly imaging exoplanets \citep{kenworthy2016high, carlomagno2020metis, brandl2018status, brandl2008metis, boccaletti2018giant, houlle2021direct}. 

Among the known ELTs with high-contrast imaging capabilities, TMT/MODHIS \citep{mawet2022fiber} was not selected due to its high spatial resolution ($\rm{R}\sim100,000$) while GMT/GMTIFS \citep{sharp2016gmtifs}  and ELT/MICADO \citep{davies2010micado} were not selected because their operating wavelengths (GMT/GMTIFS: 0.9-2.5 $\mu$m, ELT/MICADO: 0.8-2.45 $\mu$m) are similar but narrower compared to ELT/HARMONI (0.47-2.45 $\mu$m).

\section{Method} \label{sec:style_meth}

\subsection{Simulating ELT Direct Imaging of Rocky Exoplanets} \label{sec:style_ELT}

The flow chart in Figure \ref{fig:flowchart} delineates the procedure and system-related inputs utilized in the simulation. Specifically, we perform an end-to-end calculation for the photon count at an ELT instrument. For the reflection light, the photon originates from the host star, is reflected by the planet and encoded by the planet's atmosphere, then passes through the Earth's atmosphere, telescope, and a high-contrast instrument, before being recorded on a detector. For thermal emission, the only difference is that the photon originates from the planet.  

For the spectral resolution of ELT/METIS, we assume it to be 1000, which is consistent with the information provided by ESO \footnote{\url{https://elt.eso.org/instrument/METIS/}}. We assume an instrument contrast of 10$^{-4}$ for ELT/METIS, which is approximately the average post-processing contrast for ELT/METIS within 0--4.7 $\lambda/D$\citep{2016SPIE.9909E..73C}. For the instrument throughput $\eta$, we assume it to be 0.1, this is a conservative assumption, because it was assumed to be 0.36 in \citet{2020JATIS...6c5005C}.

For the resolution of ELT/HARMONI, we choose to use $R=1000$ to compare to the ELT/METIS results, although the minimum resolution of ELT/HARMONI is 3500 \citep{thatte2016elt}. While a slightly higher resolution would boost the detection $S/N$, the change from $R=1000$ to $R=3500$ is insignificant as shown in Fig. 16. in \citet{wang2017observing}.  As ELT/HARMONI exhibits a raw contrast of 10$^{-1}$ within low separation angles (0--5 $\lambda$/D), we assume a post-processing instrument contrast of 10$^{-3}$ for ELT/HARMONI, which is consistent with the extent of contrast improvement achieved using post-processing techniques with ELT/METIS \citep{2016SPIE.9909E..73C}. As with ELT/METIS, we assume an instrument throughput $\eta$ of 0.1 for ELT/HARMONI. Below we detail how each of the steps in flow chart is being simulated.

\subsubsection{Simulating Exoplanet Spectra} \label{sec:style_ELT_light_planet}

Upon identifying suitable planets, we proceed to search for appropriate atmospheric models for the candidate planets. We assume all candidate planets possess a modern Earth atmosphere with biosignatures: O$_2$, CO$_2$, CH$_4$, H$_2$O. The molecular pairs O$_2$+CH$_4$ are widely considered as biosignatures in the atmosphere of
modern Earth \citep{stueeken2020mission, crouse2021biosignature}. CH$_4$+CO$_2$ are biosignatures in the atmosphere of
Archean Earth; considering their high detectability \citep{wogan2020abundant}, we also use them as biosignatures for this project. H$_2$O is generally considered as a prerequisite for sustaining life.

We use BT-Settl high-resolution synthetic spectra \citep{allard2013bt} to simulate an M star, based on the stellar effective temperature, metallicity, and surface gravity. Assuming a star radius, $R_\star$, we can obtain the incident flux of the star ($F_\star$, unit:W m$^{-2}$) at distance $D$ with Eq. \ref{eq:general1_2}:
\begin{equation}
F_{\star} = 10^{F_0(\lambda)-8} \cdot d\lambda \cdot \frac{R_{\star}^2}{D^2 }
\label{eq:general1_2}
\end{equation}
where $F_0(\lambda)$ is the spectral template's flux array, a dimensionless quantity. $d\lambda$ is the wavelength per resolution element, computed as $\lambda_0/R$, where $\lambda_0$ is the central wavelength, $R$ is the spectral resolution.

\begin{figure*}[t]
\plotone{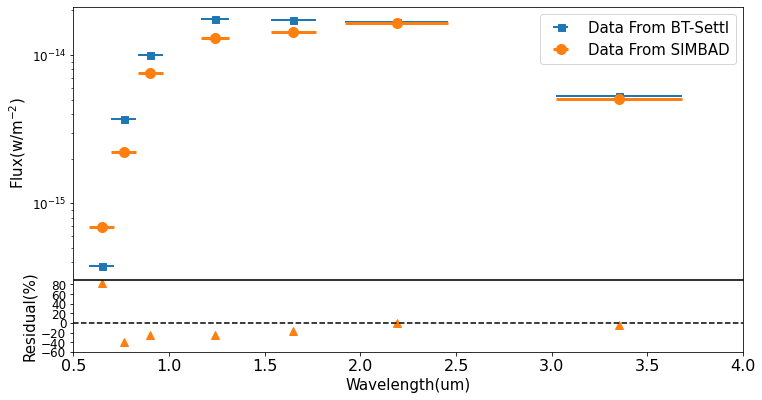}
\caption{ Top: Flux of TRAPPIST-1 from SIMBAD vs. flux of TRAPPIST-1 from BT-Settl. Bottom: the residual between SIMBAD and BT-Settl in percent. For comparison, we choose data from the $R$, $i$, $Z$, $J$, $H$, $K$, $W_1$ bands in the SIMBAD.
\label{fig:generalchecktra1}}
\end{figure*}

We use SIMBAD \citep{wenger2000simbad} photometry to check the validity of the BT-Settl spectra. Using TRAPPIST-1 as an example, we obtain $R$, $i$, $Z$, $J$, $H$, $K$, $W_1$ band photometric data from SIMBAD (shown in Figure \ref{fig:generalchecktra1}).

We calculate the incident flux of a BT-Settl spectrum with stellar parameters that best resemble TRAPPIST-1 (Table \ref{table:general1_1_transposed}). The comparison results show that the flux difference between SIMBAD photometry and a BT-Settl spectrum can be as high as 83\% but is predominantly within 25\% for the 1 to 5 micron range for which we are mainly interested and the majority of planet signal comes from. We also selected 5 host stars (with T$_{eff}$ from 2600 to 3700 K) of suitable planets to examine the difference between models and observations. The flux differences are all within 25\% for the 1 to 5 $\mu$m range.

The planet flux ($F_p$) on Earth has two components, the reflection component ($F_r$) and the thermal emission component ($F_t$). $F_r$ can be calculated via Eq. \ref{eq:general3_2}:
\begin{equation}
F_{r} = F_\star \cdot A_{g}\frac{R_{p}^2}{4a^2}
\label{eq:general3_2}
\end{equation}
where $F_\star$ is the stellar flux irradiating the planet, $A_{g}$ is wavelength-dependent albedo, as shown in Fig 8 in \citet{wang2017observing}. We employ $A_g$ to simulate the reflected spectrum by atmosphere of planet.

The albedo spectra ($A_g$) for modern Earth used in our study were derived from a high-resolution model developed by \citet{2012ApJ...752....7H, 2012ApJ...761..166H, 2013ApJ...769....6H, 2014ApJ...784...63H}. This model calculates molecular abundances across different altitudes, incorporating the effects of photochemical and disequilibrium chemistry, with details in \citet{2012ApJ...752....7H}. Cloud impacts were included by averaging clear and high-cloud scenarios, resulting in a continuum albedo of ~0.3, similar to the approach of \citet{2002AsBio...2..153D}. The spectra, covering a wavelength range of 0.5-5.6 $\mu$m with a spectral resolution of R = 500,000, include opacities of major atmospheric components and are presented as scaled albedo values.

The thermal emission component $F_t$ is calculated as follows:
\begin{equation}
F_{t} = F_{b} \cdot \bar{A_g} \cdot \theta
\label{eq:general3.5_2}
\end{equation}
where $F_{b}$ is the blackbody radiation of the planet, $\bar{A_g}$ is the normalized planet albedo spectrum, and $\theta$ is the extended solid angle of the planet as seen from Earth, which is $\frac{\pi R_p^2}{D^2}$. 

\subsubsection{Simulating Observations of ELT} \label{sec:style_ELT_obser_elt}

After arriving on Earth, the light is encoded by the Earth's atmosphere, processed by the telescope and coronagraph, and recorded by a detector. We can calculate the flux received by the detector ($F_{det}$) via Eq. \ref{eq:generaldet}:

\begin{equation}
F_{det} = \eta \cdot (F_p \cdot f_{tran} + F_\star \cdot f_{tran} \cdot C + F_{sky})
\label{eq:generaldet}
\end{equation}

where $F_p$ is the flux of the planet including its reflection and thermal components, $F_\star$ is the flux of the star on Earth, $C$ is the instrument contrast (or level of starlight suppression), $f_{tran}$ is the transmission of the sky, and $F_{sky}$ is the emission flux of the sky. We use the ESO SkyCalc Tool \citep{noll2012atmospheric, jones2013advanced} to model sky transmission and emission with an observatory altitude of 3060 m and an airmass of 1.5 for the model of sky. In calculating sky emission, we use a wavelength of 1.5 $\mu$m for ELT/HARMONI and 4.3 $\mu$m for ELT/METIS, which is the central wavelength for the two instruments in this project.

The number of photons ($N$) recorded by the detector in each wavelength channel is calculated as
follows:

\begin{equation}
N = \frac{F_{det}\cdot S\cdot t}{ E}
\label{eq:general4_2}
\end{equation}

$F_{det}$ is the flux received by the detector, $S$ is the area of the telescope for which we use 978 m$^2$, $t$ is the exposure time, and $E$ is the photon energy at central wavelength of each spectral bin.

\subsubsection{Reduced Spectrum} \label{sec:style_ELT_rem}

By the above steps, we can simulate the number of photons received by the detector during direct imaging of exoplanets. The separation of planet photons ($N_p$) from the photons received by detector ($N_{det}$) can be achieved through additional observations. 

$N_{det}$ consist of three sources: (1) photons from the star encoded by the atmosphere and the coronagraph: $N_{\star}\cdot C \cdot f_{tran}$; (2) photons from sky emission: $N_{sky}$; (3) photons from the planet encoded by the atmosphere $N_{p}\cdot f_{tran}$. 

The process of $N_p$ separation can be described as follows: (1) $N_p^\prime = N_{det} - (N_{\star}\cdot C \cdot  f_{tran} + N_{sky})$; (2) $N_p^\prime / f_{tran}$, where $N_p^\prime$ = $N_p \cdot f_{tran}$

$N_{\star}\cdot C \cdot f_{tran} + N_{sky}$ can be obtained through observations at locations where no planet signals are received. The transmission of the sky ($f_{tran}$) can be obtained through observations of a telluric standard star \citep{wang2017observing}.

\subsubsection{Noise} \label{sec:style_ELT_noise}

Based on the propagation uncertainty, we simulated the observational noise $\sigma$ after separation as follows: 

\begin{equation}
\sigma = \sqrt{\bigg(\frac{N_p\cdot \sigma_{f_{tran}}}{f_{tran}} \bigg)^2  + \frac{N_{det}+N_{\star} \cdot C \cdot f_{tran} +N_{sky}}{f_{tran}^2} }
\label{eq:general4.5_2}
\end{equation}

where $N_{\star} \cdot C \cdot f_{tran}$ is the number of stellar photons received at the telescope. $\sigma_{tran}$ is the uncertainty of sky transmission; here we assumed that $\sigma_{f_{tran}}=0.001 \cdot f_{tran}$. The derivation of Eq. \ref{eq:general4.5_2} is described in Appendix \ref{app_a}. We also assume a different fractional error at  $\sigma_{f_{tran}}=0.01 \cdot f_{tran}$ and the results are reported in Table \ref{table:selected_columns_rounded_app} and discussed in Section \ref{sec:style5.4}.

\subsection{Quantifying the Ability of ELT to Detect Biosignatures} \label{sec:style_ELT_quantity}

After establishing the method for simulating ELT observation and various noise sources, we propose a method for quantifying the $S/N$ for explanatory atmospheric gases. This $S/N$ is quantified by the $S/N$ of the gases in its observable feature wavelengths from minimum working wavelength to $\lambda_{max}$, where $\lambda_{max}$ is calculated by equating the angular resolution ($\lambda$/D) to planet-star separation. Above $\lambda_{max}$, the angular resolution falls below the planet-star separation. Therefore, we consider the planet uncharacterizable above  $\lambda_{max}$. Below we detail how each of the steps is being calculated or simulated.

\subsubsection{Signal-to-noise Ratio} \label{sec:style_snr}
\begin{figure}[t]
\includegraphics[width=8.5cm]{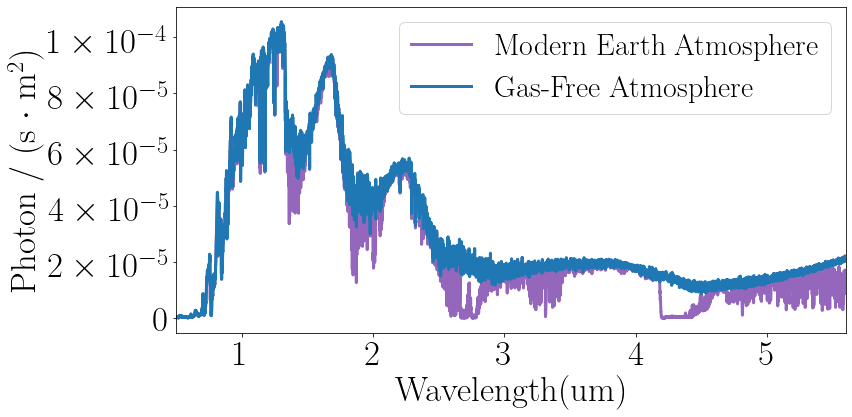}
\caption{TRAPPIST-1~e photon flux detected on Earth, assuming TRAPPIST-1~e has a biosignature-free and modern Earth atmosphere. The difference between the two curves is caused by biosignature gases. The greater the differences, the stronger the impact of gases on the observed spectrum, resulting in a higher $S/N$ of gases. We assumed a spectral resolution of 1000.
\label{fig:general2_3}}
\end{figure}

\begin{figure*}[t]
\plotone{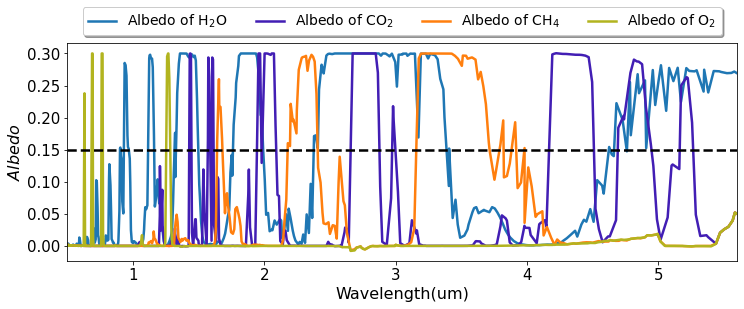}
\caption{ The albedo of H$_2$O, CO$_2$, CH$_4$, O$_2$ in the atmosphere of modern Earth. The interval where the albedo of a gas exceeds 0.15 is considered as the feature wavelength for that gas. The black dashed line is the criteria for albedo of 0.15. The wavelengths where two or more gases have overlapping features will not be considered as feature wavelengths. The wavelengths of the intervals with albedo higher than 0.15 are described in Table \ref{table:feature}.
\label{fig:generalfeature}}
\end{figure*}

\begin{table}[]
\setlength{\arrayrulewidth}{0.4mm}
\begin{tabular}{c c c c} 
 \hline\hline
\multicolumn{4}{c}{Feature Wavelength($\mu$m)} \\ [0.5ex] 
 \hline
 H$_2$O & CH$_4$ & CO$_2$ & O$_2$ \\ 
 \hline
 0.90-0.91 & 1.64-1.68 & 1.57-1.59 & 0.62-0.63\\
 
 0.93-0.97 & 2.18-2.38 & 1.60-1.61 & 0.68-0.69\\
  
 1.11-1.16 & 3.88-3.41 & 1.99-2.09 & 0.76-0.77\\
   
 1.31-1.32 & 3.41-3.71 & 4.17-4.51 & 1.25-1.28\\
 
 1.33-1.43 & 3.79-3.82 & - & -\\
 
 1.45-1.50 & 3.88-3.92 & - & -\\
 1.76-1.94 & 3.98-3.99 & - & -\\
 1.97-1.98 & - & - & -\\
 2.41-2.45 & - & - & -\\
 3.00-3.16 & - & - & -\\
 4.62-4.63 & - & - & -\\
 4.66-4.69 & - & - & -\\
 4.95-5.16 & - & - & -\\
 5.27-5.60 & - & - & -\\[1ex] 
 \hline
\end{tabular}
\caption{Feature wavelengths of biosignature gas we set for modern Earth atmosphere.
\label{table:feature}}
\end{table}

We define $S/N$ for the gas in the atmospheres of exoplanets with Eq. \ref{eq:single_feature_3}:

\begin{equation}
\label{eq:single_feature_3}
S/N  =  \frac{ N' - N}{\sigma}
\end{equation}

where $N$ is the observed photon number for the planet. $N$ and $\sigma$ can be obtained through ELT observational procedure as shown in \S \ref{sec:style_ELT}. $N'$ is the observed photon number for the planet, assuming the planet has a biosignature-free (no CO$_2$, CH$_4$, H$_2$O, CO) modern Earth atmosphere. $N'$ can be obtained through the same procedure as for $N$, with the only difference being that $N'$ assumes the planet with a modern Earth atmosphere that lacks CH$_4$, CO$_2$, O$_2$, H$_2$O. 

To obtain a biosignature-free modern Earth atmosphere, we remove the absorption line of the CH$_4$, CO$_2$, O$_2$, H$_2$O by finding the envelope of the albedo of modern Earth. To find the envelope, we divided the spectrum into 430 intervals, using the largest value in each interval to construct the envelop. Choosing TRAPPIST-1~e (archetype of nearby transiting terrestrial  planet) as an example, the spectrum of a biosignature-free atmosphere and the modern Earth atmosphere are shown in Figure \ref{fig:general2_3}.

\subsubsection{Quantifying $S/N$ of ELT Direct Imaging Planet Atmosphere} \label{sec:style_DetectionCapabilit}

\begin{table*}[t]
\setlength{\arrayrulewidth}{0.4mm}
\begin{tabular}{c c c c c c c c} 
 \hline\hline
\multicolumn{8}{c}{ELT/METIS} \\ [0.5ex] 
 \hline
Rank & Planet & $S/N_{total}$ & $S/N_{H_2O}$ & $S/N_{CH_4}$ & $S/N_{CO_2}$ & $S/N_{O_2}$ &
T$_{S/N=5}$(h)\\ 
 \hline
1 & GJ 887 b & 18.24 & 4.81 & 17.59 & 0.00 & 0.0 & 0.75 \\
2 & Proxima b & 2.22 & 1.57 & 1.16 & 1.05 & 0.0 & 51 \\
3 & Wolf 1061 c & 0.72 & 0.24 & 0.67 & 0.00 & 0.0 & 490 \\
4 & GJ 682 b & 0.13 & 0.13 & 0.00 & 0.00 & 0.0 & 14300 \\
5 & GJ 667 C c & 0.07 & 0.07 & 0.00 & 0.00 & 0.0 & 54500 \\[1ex]
 \hline \hline
\multicolumn{8}{c}{ELT/HARMONI} \\ [0.5ex] 
 \hline
Rank & Planet &$S/N_{total}$ & $S/N_{H_2O}$ & $S/N_{CH_4}$ & $S/N_{CO_2}$ &  $S/N_{O_2}$ &
T$_{S/N=5}$(h)\\ 
 \hline
1 & GJ 887 b & 4.94 & 3.18 & 2.74 & 2.40 & 1.02 & 10.5 \\
2 & Wolf 1061 b & 1.04 & 0.94 & 0.12 & 0.20 & 0.37 & 235 \\
3 & Wolf 1061 c & 0.64 & 0.43 & 0.34 & 0.29 & 0.13 & 625 \\
4 & Proxima b & 0.49 & 0.39 & 0.23 & 0.19 & 0.09 & 1050 \\
5 & GJ 682 b & 0.47 & 0.36 & 0.23 & 0.19 & 0.09 & 1150 \\
6 & GJ 1061 c & 0.44 & 0.41 & 0.09 & 0.06 & 0.11 & 1300 \\
7 & GJ 667 C c & 0.26 & 0.18 & 0.14 & 0.12 & 0.06 & 3800 \\
8 & GJ 1061 d & 0.25 & 0.19 & 0.12 & 0.10 & 0.05 & 4000 \\
9 & GJ 1002 b & 0.13 & 0.12 & 0.03 & 0.02 & 0.03 & 14500 \\
10 & GJ 1002 c & 0.08 & 0.06 & 0.04 & 0.03 & 0.01 & 39000 \\ \hline
\end{tabular}
\caption{$S/N$ of biosignatures for candidate planets with ELT/METIS and ELT/HARMONI direct imaging. T$_{S/N=5}$(s): required exposure time to reach a $S/N_{total}$ of 5. We assume an exposure time of 10 hours and a resolution of 1000 for ELT, a coronagraph contrast of 10$^{-3}$ for ELT/HARMONI and 10$^{-4}$ for ELT/METIS.}
\label{table:selected_columns_rounded_5}
\end{table*}

Building on the work of \citet{2021ApJ...923..144P}, we have developed a method to quantitatively evaluate the $S/N$ of ELT for planetary atmospheric gases with Eq. \ref{eq:muti_feature_3}:

\begin{equation}
\label{eq:muti_feature_3}
S/N_{gas}  =  \sqrt{\sum\limits_{i}S/N_{i}^2}
\end{equation}

where $S/N_{gas}$ is the signal-to-noise ratio of gas in Eq. \ref{eq:single_feature_3}, and $i$ is the index for each feature wavelength as shown in Table \ref{table:feature}.

We identify the feature wavelengths based on the albedo spectrum of modern Earth. The region where the albedo of a gas is all above 0.15 (half the albedo of the modern Earth) is considered to be the feature wavelength of that gas. The wavelengths where two or more gases have overlapping features will not be considered. Based on the assumed Earth atmosphere model ($A_g$), the feature wavelengths of each gas have been determined. These feature wavelengths are described in Table \ref{table:feature}, shown in Figure \ref{fig:generalfeature}.

For example, when calculating the $S/N$ for ELT/HARMONI observations of Wolf 1061 b, the observable wavelengths are from 0.45 $\mu$m (minimum working wavelength for ELT/METIS) to $\lambda_{max}$-1.66 $\mu$m (as shown in Table \ref{table:general1_4}). Feature wavelengths of CO$_2$ are therefore 1.57-1.59 $\mu$m and 1.60-1.61 $\mu$m. The $S/N$ of CO$_2$ can be quantified by summing up all the $S/N$ above two CO$_2$ wavelength regions using Eq. \ref{eq:muti_feature_3}. The $S/N$ of the atmosphere will also be calculated through a similar process, with the difference being that the feature wavelength will be the merged feature wavelengths for all gases in the atmosphere up to $\lambda_{max}$.

\section{Results} \label{sec:styleApplication}

We applied the above method to the candidate planets and obtained observational results in the wavelength ranges of 0.5-2.45 $\mu$m for ELT/HARMONI and 3-5.6 $\mu$m for ELT/METIS with 10 hours exposure times. Based on the observational results, we quantified and ranked the $S/N$ of ELT/METIS and ELT/HARMONI for biosignatures in the atmosphere of the candidate planets, as shown in Table \ref{table:selected_columns_rounded_5}. The ranking indicates that GJ 887 b is the most suitable known planet for direct imaging of its atmosphere with ELT/METIS and ELT/HARMONI, and Proxima b is the most suitable known planet for direct imaging of CO$_2$ with ELT/METIS.

Considering the diversity of instrument contrasts, we explored the $S/N$ of ELT/METIS and ELT/HARMONI in the case of different instrument contrast (for ELT/METIS: 10$^{-3}$ to 10$^{-6}$; for ELT/HARMONI: 10$^{-2}$ to 10$^{-5}$;) for the highest ranked target, GJ 887~b, as shown in Table \ref{table:general6_4}. We conclude that the $S/N$ of ELT/METIS and ELT/HARMONI will improve with the enhancement of the instrument contrast.  Specifically, the enhancement of instrument contrast contributes most significantly to the improvement in the $S/N$ of ELT/HARMONI. We detail the results below.

\begin{table}[b]
\setlength{\arrayrulewidth}{0.4mm}
\begin{tabular}{c c c c c c} 
 \hline\hline
\multicolumn{6}{c}{$S/N$ of ELT/METIS for GJ 887~b} \\ [0.5ex] 
 \hline
C &H$_2$O & CH$_4$ & CO$_2$ &  O$_2$ & Total \\ 
 \hline
10$^{-3}$& 1.92 & 7.05 & 0 & 0 & 7.31 \\

 $^*$10$^{-4}$&  4.81 &  17.59 &  0 &  0 &  18.24 \\

10$^{-5}$& 8.10 & 30.11 & 0 & 0 & 31.18 \\

10$^{-6}$& 9.58 & 36.57 & 0 & 0 & 37.81 \\[1ex]
 \hline \hline
\multicolumn{6}{c}{$S/N$ of ELT/HARMONI for GJ 887~b} \\ [0.5ex] 
 \hline
C &H$_2$O & CH$_4$ & CO$_2$ &  O$_2$ & Total \\ 
 \hline
10$^{-2}$& 1.00 & 0.86 & 0.76 & 0.32 & 1.56 \\

 $^*$10$^{-3}$& 3.18 & 2.74 & 2.40 & 1.02 & 4.94 \\

10$^{-4}$& 10.05 & 8.66 & 7.60 & 3.22 & 15.62 \\

10$^{-5}$& 31.67 & 27.31 & 23.99 & 10.16 & 49.29 \\[1ex]
 \hline
\end{tabular}
\caption{$S/N$ of ELT/METIS and ELT/HARMONI for GJ 887~b. C: Instrument contrast. The rows with $^*$ are the data for the assumed contrast of this project.
\label{table:general6_4}}
\end{table}

\subsection{ELT/METIS} \label{sec:styleMETIS}
We have quantified the $S/N$ of ELT/METIS for direct imaging of exoplanet atmospheres within 3-5.6 $\mu$m. Based on the quantitative results, we present a ranking of candidate planets in terms of their $S/N$. According to the ranking, we find that the three most detectable planets are GJ 887 b, Proxima b, and Wolf 1061 c, while the least detectable planet is GJ 667 C c. GJ 667 C c requires an exposure time approximately 54,500 hours in order to achieve $S/N = 5$. With a 10 hours exposure time, the $S/N$ of GJ 887 b far surpasses that of other planets, being approximately 8.2 times higher than that of Proxima b, 25.3 times higher than that of Wolf 1061 c, and approximately 140-260 times higher than the remaining planets.

The high $S/N$ of GJ 887 b is reasonable due to its abundant reflection flux and thermal flux contributions. The abundant reflection flux and thermal flux in the planet are attributed to the high luminosity of its host star \citep[the brightest red dwarf,][]{jeffers2020multiplanet}, as well as its small semi-major axis and large planetary radius. These conditions allow it to have a high equilibrium temperature and reflect a significant amount of light from the star. 

In terms of gas $S/N$ (or $S/N_{gas}$), we find that ELT/METIS has the capability to detect CO$_2$, CH$_4$, and H$_2$O in the NIR band, but lacks the capability to detect O$_2$. At an exposure time of 10 hours, for planets with $S/N >$ 5 (GJ 887 b), CH$_4$ exhibits the highest $S/N$, approximately 4 times stronger than that of H$_2$O, but CO$_2$ and O$_2$ are not detectable due to the observable wavelength (3-3.94 $\mu$m). For other planets, CO$_2$ is only detectable for Proxima b, and O$_2$ is not detectable for any of them, as the observable wavelengths do not include the feature wavelengths of CO$_2$ and O$_2$.

CO$_2$ is only observable for Proxima b with ELT/METIS due to its large angular separation. The angular separation of Proxima b allows its observable wavelengths for ELT/METIS to range from 3 to 7.11 $\mu$m (see $\lambda_{max}$ in Table \ref{table:general1_4}), which includes strong features of CO$_2$ from 4.17 to 4.51 $\mu$m (see Table \ref{table:feature}). In contrast, the observable wavelengths for other planets with ELT/METIS are limited to within 4 $\mu$m, which do not cover the feature wavelengths of CO$_2$.

For the impact of instrument contrast on the $S/N$ of the ELT/METIS, from $C=10^{-3}$ to 10$^{-4}$, 10$^{-4}$ to 10$^{-5}$, and 10$^{-5}$ to 10$^{-6}$, the $S/N$ of ELT/METIS for GJ 887 b improves by 149.5$\%$, 70.9$\%$ and 21.3$\%$, respectively. Once $C$ reaches 10$^{-5}$, the enhancement of instrument contrast provides less aid to the performance of ELT/METIS. This is because the dominating noise source changes from stellar noise to sky background noise at higher contrast levels.

The range of contrast is chosen to be consistent with the expected contrast levels achievable with multiple post-processing techniques at different angular separations \citep{2016SPIE.9909E..73C}.

\subsection{ELT/HARMONI} \label{sec:styleHARMONI}

We have quantified $S/N$ of ELT/HARMONI within 0.5-2.45 $\mu$m. Based on the quantitative results, GJ 887 b remains the most detectable planet, while GJ 1002 c is the least detectable planet. For the GJ 887 b (the planet with highest $S/N$),  ELT/HARMONI requires approximately 14 times longer exposure times ($\sim$10.5 h) than ELT/METIS (0.75 h) to achieve $S/N=5$ (see Table \ref{table:selected_columns_rounded_5}).

For other planets,  the $S/N$ of ELT/HARMONI is estimated to be approximately 0.22 to 3.71 times that of ELT/METIS, for an exposure time of 10 hours.

Our results demonstrate that ELT/HARMONI possesses $S/N$ for all four biosignatures (H$_2$O, CO$_2$, CH$_4$, O$_2$). The $S/N$ of H$_2$O is the highest, while the $S/N$ of O$_2$ is the lowest.

For different instrument contrast, from $C=10^{-2}$ to $C=10^{-3}$, $10^{-3}$ to $10^{-4}$, $10^{-4}$ to $10^{-5}$, the $S/N$ of ELT/HARMONI for GJ 887 b improves by 216.7$\%$, 216.2$\%$ and 215.6$\%$, respectively. The enhancement of instrument contrast will greatly improve the $S/N$ of ELT/HARMONI. In the wavelength (0.47-2.45 $\mu$m) relevant to ELT/HARMONI, the noise is mainly from the host star. Such noise can be mitigated by improving the performance of the coronagraph. The setting of the contrast range is akin to that for ELT/METIS, but with a different baseline contrast (from $10^{-4}$ to $10^{-3}$, see Table \ref{table:general6_4}).

\section{Discussion} \label{sec:styledisc}

\subsection{ELT vs JWST}\label{sec:style5.1}

\begin{table*} \setlength{\arrayrulewidth}{0.4mm} \begin{tabular}{c c c c c c c} 
\hline\hline Telescope &Planet & H$_2$O & CH$_4$ & CO$_2$ & O$_2$ & $S/N_{total}$ \\ [0.5ex] \hline 
JWST/NIRSpec &TRAPPIST-1~e & 34.85$\pm$2.02 & 20.13$\pm$2.19 & 16.86$\pm$1.01 & 5.71$\pm$0.95 & 44.04$\pm$2.80 
\\ JWST/NIRSpec &TRAPPIST-1~h & 35.30$\pm$2.16 & 21.20$\pm$1.78 & 17.30$\pm$1.23 & 5.97$\pm$1.17 & 45.09$\pm$2.85 
\\ ELT/HARMONI&TRAPPIST-1~e & 0.000 & 0.000 & 0.000 & 0.000 & 0.000 

\\ ELT/HARMONI&TRAPPIST-1~h & 0.001 & 0.000 & 0.000 & 0.000 & 0.001 \\[1ex] \hline \end{tabular} \caption{The $S/N$ of biosignatures in the atmospheres of TRAPPIST-1 e and h with ELT/HARMONI direct imaging and JWST transit spectroscopy. \label{table:general8_1_transposed}} \end{table*}

\citet{batalha2017information} noted that the launch of JWST will revolutionize our understanding of planetary atmospheres. We compare the direct imaging of ELT with JWST's transit spectroscopy. Our results indicate that for those planets with small angular separation and high contrast, JWST's transit spectroscopy may exhibit better performance compared to the direct imaging methods of ELT and JWST. Below we present these discussions in more detail.

\subsubsection{ELT Direct Imaging vs JWST Transit Spectroscopy}\label{sec:style5.2.2}

The transit spectroscopy of JWST is another method to explore exoplanet atmospheres. \citet{greene2016characterizing} highlighted its favorable performance in probing exoplanet atmospheres in 1-2.5 $\mu$m, which may compensate for the limitations of ELT direct imaging in this band.

For comparison, we simulated and quantified the $S/N$ of ELT's direct imaging mode and JWST's transit spectroscopy for the same targets - TRAPPIST-1e and h, two planets in the TRAPPIST-1 system with medium to largest angular separation among all TRAPPIST planets. TRAPPIST-1 e and h have angular separations of 2.42 mas and 5.11 mas, respectively, which are smaller than the minimum spatial resolution of ELT/METIS (15.74 mas), we therefore do not consider ELT/METIS in the TRAPPIST -1 discussion. The results are presented in Table \ref{table:general8_1_transposed}. Based on the table, we observe that JWST's transit spectroscopy exhibits superior $S/N$ for the atmospheres of high-contrast, low-separation planets like those in the TRAPPIST-1 system compared to ELT's direct imaging. For the simulation of ELT direct imaging TRAPPIST-1~e and h, we utilized the methodology describe in \S \ref{sec:style_meth}.

For JWST, we employed PICASO \citep{batalha2019exoplanet} to model the planet's transmission spectra, assuming a modern Earth atmosphere \citep{cox2015allen}. Based on the obtained transmission spectra, we simulated JWST/NIRSpec \citep[0.6-5.3 $\mu$m,][]{2007SPIE.6692E..0MB} observations using PandExo \citep{batalha2017pandexo}. For the observation strategy settings for PandExo, we set the saturation level equal to  80(in percent); the number of transits is 10; the noise floor is 0; R(Resolution)=150. The settings for PandExo were consistent with the strategy outlined in \citep{mikal2022detecting}. To make sure we used the same setting, we reproduced the results in \citet{mikal2022detecting}. Using the observational results, we quantified $S/N$ of JWST transit spectroscopy for planetary atmospheres using the method described in \citet{2021ApJ...923..144P}. The result of a detection of S/N$\sim$5 for O$_2$ in 10 transits is consistent with previous studies that concluded O$_2$ can be detected for 7-40 transits wiht JWST~\citep{Yaeger2019,Gialluca2021}.  

The parameters for TRAPPIST-1e and h are described in Table \ref{table:general1_1_transposed}. The selection of the TRAPPIST-1 system was motivated by its proximity and the presence of one of the closest known terrestrial planets in transit.

\subsection{ELT: The Performance of Different Detection Method}\label{sec:style5.2}

The ELT will greatly enhance our understanding of exoplanet atmospheres in multiple aspects. \citet{currie2023there} explored the $S/N$ of ELT's high-resolution transit spectroscopy for exoplanet atmospheres, noting that the high-resolution transit spectroscopy of ELT exhibits strong $S/N$ for O$_2$, CH$_4$, and CO$_2$ in transiting planets. High-resolution transit spectroscopy appears suitable for the detection of O$_2$ in exoplanets with transit phenomena \citep{serindag2019testing}. However, \citet{hardegree2023bioverse} pointed out that ELT's transit spectroscopy requires about 16-55 yr to detect O$_2$ on TRAPPIST-1 d-g. 

Our results indicate ELT/HARMONI's ability to detect O$_2$ through direct imaging depends on the instrument contrast. $S/N$ will increase approximately by 30 times from 0.3 to 10, when the contrast ($C$) improves from $10^{-2}$ to $10^{-5}$.

In addition to O$_2$, our findings also indicate that ELT/METIS exhibits strong $S/N$ for CH$_4$, assuming instrument contrast $C=10^{-4}$. In general, direct imaging is an important method for exploring nearby rocky planet atmospheres as it does not rely on transits which are extremely rare (with only one exception, HD 219134 b) for nearby ($<12$ pc) rocky planets.

Furthermore, the performance of ELT in the high spectral resolution direct imaging mode can be improved more for low-noise detectors, reducing the contrast requirements for planets by 2-3 orders of magnitude under high-dispersion conditions \citep{wang2017observing}.

\subsection{$S/N$ of Biosignature Pairs} \label{sec:style5.1}

The presence of gases in specific combinations is related to thermal-dynamical imbalance between them, rather than a single molecule (e.g., CO$_2$ in biosignature pair of CH$_4$ + CO$_2$). The high $S/N$ of a single molecule alone does not necessarily prove the high $S/N$ of disequilibrium phenomena. Therefore, the $S/N$ of disequilibrium pairs composed of multiple molecules is a topic worth exploring. 

We propose a quantitative approach to characterize the $S/N$ of thermal-dynamical imbalance pairs. It takes into account the number of molecules involved in the composition of the disequilibrium pair and their $S/N$, as following:

\begin{equation}
\label{eq:comb_1}
S/N_{total} = \bigg(\prod \limits_{i=1}^nS/N_{gas\ i}\bigg)^\frac{1}{n} 
\end{equation}

where $S/N_{gas\ i}$ is the $S/N$ of single gas.

As an example, according to Table \ref{table:general8_1_transposed}, TRAPPIST-1~e under JWST observation has a $S/N$ of 44.04. However, due to the low $S/N$ of O$_2$ ($S/N$=5.71), which will limit the detection of the biosignature pair of O$_2$ + CH$_4$.

\subsection{Fidelity of the Calculations} \label{sec:style5.4}

The assumptions such as instrument sensitivity, telluric uncertainty, and the usage of feature wavelengths in this paper will lead to differences in the calculations. Below we detail the impact of each assumption on our calculations. As shown below, our calculations provide a conservative estimate for $S/N$ and exposure times.

Considering telluric variability as a source of noise may impact $S/N$, we explored two cases for telluric removal residual at 0.1\% and 1\% levels. The results are reported in Table \ref{table:selected_columns_rounded_5} and \ref{table:selected_columns_rounded_app} for the two cases. The results indicate that telluric variability has a limited impact on ELT/HARMONI and ELT/METIS. This is reasonable, as its influence is only related to the planetary photon noise $(\frac{N_p\cdot \sigma_{f_{tran}}}{f_{tran}})$ in the noise calculations (Eq. \ref{eq:general4.5_2}), and the planetary photon noise does not dominate the noise budget.

In terms of the impact of instrumental curves (instrument throughput and coronagraph contrast) on the calculations, generally a higher throughput and contrast results in a higher $S/N$. The improvement in $S/N$ due to increased throughput roughly follows a square root law. We have assumed an instrumental throughput of 0.1 for both ELT/HARMONI and MELT/ETIS, which is a conservative assumption. The actual throughput varies with different observation modes and working wavelengths, but it is generally higher than 0.1~\citep{2022SPIE12185E..5LC}. 

The gains in $S/N$ due to improvements in coronagraph contrast diminish gradually as the thermal emission begins to dominate the noise budget. For instance, the working wavelength of ELT/METIS determines that it is more affected by atmospheric thermal noise, therefore the improvement effect on $S/N$ diminishes progressively from contrasts of 10$^{-3}$ to 10$^{-6}$, and only 21.3\% enhancement from 10$^{-5}$ to 10$^{-6}$. ELT/HARMONI, with working wavelengths of 0.45-2.5 $\mu$m, avoids the high atmospheric thermal noise regions and demonstrates an increase of over 200\% in $S/N$ across all four explored cases from $10^{-2}$ to $10^{-5}$ (see Table \ref{table:general6_4}).

Another source of uncertainty is the selection of gas feature wavelengths, we have ignored regions where the gas albedo is below 0.15, leading to an underestimation of the gas $S/N$. However, the method we employed still includes the majority of absorption features of the gases. The ignored regions constitute approximately 23\% of the total absorption signal. The weak and blended features that are ignored in this work can be better recovered by spectral retrievals \citep[e.g., ][]{2023arXiv231009902R,2023AJ....166..203W}. This will be the main focus of the next phase of this project.

\section{Summary} \label{sec:stylesum}

We evaluated 10 and 5 rocky planets with ELT/HARMONI and ELT/METIS, respectively. We simulated high-contrast, medium-resolution ($R=1000$) direct imaging of these planets with ELT/METIS and ELT/HARMONI in 3-5.6 $\mu$m and 0.5-2.45 $\mu$m. We investigated and quantified the $S/N$ of ELT/METIS and ELT/HARMONI for the atmospheres and potential biosignatures therein. We drew conclusions with ELT and compared them with JWST. Here are our primary findings:
\begin{itemize}
  \item  We find that the direct imaging mode of ELT/METIS has the capability to detect CH$_4$, CO$_2$, and H$_2$O in 3-5.6 $\mu$m. The direct imaging mode of ELT/HARMONI has the ability to detect CH$_4$, CO$_2$, O$_2$, and H$_2$O, but it requires more exposure time, especially for O$_2$. The direct imaging mode of ELT/METIS performs better for three planets (GJ 887 b, Proxima b, and Wolf 1061 c) with the higher $S/N$ than ELT/HARMONI. ELT/HARMONI performs better for planets other than GJ 887 b, Proxima b, and Wolf 1061 c (See \S\ref{sec:styleMETIS}\&\ref{sec:styleHARMONI}).  
  \item We investigate the impact of coronagraph performance on the $S/N$. We find that once the instrument contrast $C$ exceeds 10$^{-5}$, improvement in the coronagraph performance is less helpful for the $S/N$ of ELT/METIS because the thermal emission begins to dominate the noise budget. From $C=10^{-2}$ to 10$^{-5}$, the improvement in performance of the coronagraph greatly aids in the $S/N$ of ELT/HARMONI (See \S\ref{sec:styleMETIS}\&\S \ref{sec:styleHARMONI}).
  \item GJ 887 b is currently the most amenable targets for direct imaging with ELT, as it has a bright host star, close orbit, large size, which result in the highest $S/N$ (See \S \ref{sec:styleMETIS}\&\S \ref{sec:styleHARMONI}, and Table \ref{table:selected_columns_rounded_5}).
  \item Proxima Cen b is a unique target for direct imaging with ELT/METIS, not only because it can achieve a $S/N$ of 5 within 51 hours but also because its angular separation enables the direct observation of CO$_2$, H$_2$O, and CH$_4$ (See \S \ref{sec:styleMETIS} \& Table \ref{table:selected_columns_rounded_5}).
  \item We find that for transiting planets with high contrast and small separation, such as the TRAPPIST-1 system, JWST transit spectroscopy is more suitable than ELT direct imaging in terms of the achievable detection $S/N$ (See \S\ref{sec:styledisc}). 
\end{itemize}

\section*{Acknowledgements}

\par We thank the anonymous referee for providing comments and suggestions that significantly improve the manuscript. H.Z. acknowledges the support by the Undergraduate Research Apprenticeship Program from the Office of Academic Enrichment of The Ohio State University. M.K.P would like to thank the United States Air Force Academy, Department of Physics and Meteorology for sponsoring the graduate work of the M.K.P. Additionally, we want to acknowledge the hard work and expertise of the scientific, technical, and administrative staff at the Large Binocular Telescope. J.W. acknowledges the support by the National Science Foundation under Grant No. 2143400. We would like to thank the Group for Studies of Exoplanets (GFORSE) at The Ohio State University, Department of Astronomy for continuous feedback throughout the development of this research. This research has made use of the NASA Exoplanet Archive, which is operated by the California Institute of Technology, under contract with the National Aeronautics and Space Administration under the Exoplanet Exploration Program.

\par The views expressed in this article are those of the author and do not necessarily reflect the official policy or position of the Air Force, the Department of Defense, or the U.S. Government.

\clearpage

\appendix

\section{Derivation of Equation 8}\label{app_a}

The photons received at the detector ($N_{det}$) consist of stellar photons ($N_{s}$) and planetary photons ($N_{p}$) transmitted through Earth's atmosphere ($f_{tran}$) and instrument's coronagraph ($C$) as well as photons from the background sky radiance ($N_{sky}$): 

\begin{equation}
N_{det} = N_p \cdot f_{tran} + N_s \cdot f_{tran}  \cdot C + N_{sky}
\end{equation}

with a photon noise of $\sqrt{N_{det}}$:

\begin{equation}
\sigma_{N_{det}} = \sqrt{N_p \cdot f_{tran} + N_s \cdot f_{tran}  \cdot C + N_{sky}}
\end{equation}

To find $N_{p}$, we could remove the stellar and background sky photons through additional observations ($N_{add}$) which consists of stellar photon ($N_{s}$) processed by the sky ($f_{tran}$) and coronagraph ($C$) as well as photons from the sky ($N_{sky}$), with a photon noise of $\sqrt{N_{add}}$. 

that is:

\begin{equation}
N_{add} =  N_s \cdot f_{tran}  \cdot C + N_{sky}
\end{equation}

with photon noise:

\begin{equation}
\sigma_{N_{add}} = \sqrt{ N_s \cdot f_{tran}  \cdot C + N_{sky}}
\end{equation}

We can compute the planetary signal transmitted through the atmosphere ($N_{p}\cdot f_{tran}$) by subtracting $N_{add}$ from $N_{det}$:

\begin{equation}
N_{p} \cdot f_{tran} =  N_{det} - N_{add} = \Delta N
\end{equation}

with photon noise:

\begin{equation}
\sigma_{N_{p} \cdot f_{tran}} = \sqrt{\sigma_{N_{det}}^2 + \sigma_{N_{add}}^2} = \sigma_{\Delta N}
\end{equation}.

We remove the sky transmission ($f_{tran}$) through observations of a telluric standard star and divide by this signal:

\begin{equation}
N_{p}  =  \frac{N_{det} - N_{add}}{f_{tran}} = \frac{\Delta N}{f_{tran}}
\end{equation}

with photon noise \citep[propagation of uncertainty,][]{1997ieas.book.....T}:

\begin{equation}
\frac{\sigma_{N_{p}}^2}{N_{p}^2} = \frac{\sigma_{\Delta N}^2}{\Delta N^2} + \frac{\sigma_{f_{tran}}^2}{f_{tran}^2}
\end{equation}

so the final noise ($\sigma_{N_p}$) is:

\begin{equation}
\sigma_{N_{p}} = \sqrt{\frac{\sigma_{\Delta N}^2}{\Delta N^2} \cdot N_p^2 + \frac{\sigma_{f_{tran}}^2}{f_{tran}^2} \cdot N_p^2}
\end{equation}

note that $\Delta N = N_p \cdot f_{tran}$, and $\sigma_{f_{tran}}=0.001 \cdot f_{tran}$, which implies:

\begin{equation}
\sigma_{N_{p}} = \sqrt{\frac{\sigma_{\Delta N}^2}{f_{tran}^2} + \bigg(\frac{N_p\cdot \sigma_{f_{tran}}}{f_{tran}}\bigg)^2}
\end{equation}

and $\sigma_{\Delta N}=\sqrt{\sigma_{N_{det}}^2 + \sigma_{N_{add}}^2}=\sqrt{N_{det}+ N_{add}}$, which means:

\begin{equation}
\sigma_{N_{p}} = \sqrt{\frac{N_{det}+N_{add}}{f_{tran}^2} + \bigg(\frac{N_p\cdot \sigma_{f_{tran}}}{f_{tran}}\bigg)^2}
\end{equation}

or

\begin{equation}
\sigma_{N_{p}} = \sqrt{\bigg(\frac{N_p\cdot \sigma_{f_{tran}}}{f_{tran}}\bigg)^2  + \frac{N_{det}+N_{\star} \cdot C \cdot f_{tran} +N_{sky}}{f_{tran}^2} }
\label{eq:general4.5_2_app}
\end{equation}

\clearpage
\section{additional case}\label{app_b}

\begin{table*}[h]
\setlength{\arrayrulewidth}{0.4mm}
\begin{tabular}{c c c c c c c c} 
 \hline\hline
\multicolumn{8}{c}{ELT/METIS} \\ [0.5ex] 
 \hline
Rank & Planet & $S/N_{total}$ & $S/N_{H_2O}$ & $S/N_{CH_4}$ & $S/N_{CO_2}$ & $S/N_{O_2}$ &
T$_{S/N=5}$(h)\\ 
 \hline
1 & GJ 887 b & 18.16 & 4.80 & 17.52 & 0.00 & 0.0 & 0.75 \\
2 & Proxima b & 2.22 & 1.57 & 1.16 & 1.05 & 0.0 & 51 \\
3 & Wolf 1061 c & 0.71 & 0.24 & 0.67 & 0.00 & 0.0 & 490 \\
4 & GJ 682 b & 0.13 & 0.13 & 0.00 & 0.00 & 0.0 & 14850 \\
5 & GJ 667 C c & 0.07 & 0.07 & 0.00 & 0.00 & 0.0 & 55000 \\\hline
\hline
\multicolumn{8}{c}{ELT/HARMONI} \\ [0.5ex] 
 \hline
Rank & $S/N$ &$S/N_{total}$ & $S/N_{H_2O}$ & $S/N_{CH_4}$ & $S/N_{CO_2}$ &  $S/N_{O_2}$ &
T$_{S/N=5}$(h)\\ 
 \hline
1 & GJ 887 b & 4.94 & 3.17 & 2.74 & 2.40 & 1.02 & 10.5 \\
2 & Wolf 1061 b & 1.04 & 0.94 & 0.12 & 0.20 & 0.37 & 235 \\
3 & Wolf 1061 c & 0.64 & 0.43 & 0.34 & 0.29 & 0.13 & 625 \\
4 & Proxima b & 0.49 & 0.38 & 0.23 & 0.19 & 0.09 & 1050 \\
5 & GJ 682 b & 0.47 & 0.36 & 0.23 & 0.19 & 0.09 & 1150 \\
6 & GJ 1061 c & 0.44 & 0.41 & 0.09 & 0.06 & 0.11 & 1300 \\
7 & GJ 667 C c & 0.26 & 0.17 & 0.14 & 0.12 & 0.05 & 3800 \\
8 & GJ 1061 d & 0.25 & 0.19 & 0.12 & 0.10 & 0.05 & 4000 \\
9 & GJ 1002 b & 0.13 & 0.12 & 0.03 & 0.02 & 0.03 & 14500 \\
10 & GJ 1002 c & 0.08 & 0.06 & 0.04 & 0.03 & 0.01 & 39000 \\ \hline
\end{tabular}
\caption{Biosignature $S/N$ for candidate planets with ELT/HARMONI and ELT/METIS direct imaging. T$_{S/N=5}$(s): required exposure time to reach a $S/N_{total}$ of 5. We assume an exposure time of 10 hours and a resolution of 1000 for ELT, a coronagraph contrast of 10$^{-3}$ for ELT/HARMONI and 10$^{-4}$ for ELT/METIS, and the uncertainty of sky transimssion ($\sigma_{f_{tran}}$) equals to $0.01\cdot f_{tran}$.}
\label{table:selected_columns_rounded_app}
\end{table*}

\clearpage
\bibliography{sample631}{}
\bibliographystyle{aasjournal}

\end{CJK*}
\end{document}